\documentclass[pre,
 reprint,
 amsmath,amssymb,
 aps,nofootinbib,
]{revtex4-2}

\usepackage{url}
\usepackage{amsmath,amssymb,graphicx,bbm}
\usepackage{mwe}
\usepackage{rotating}
\usepackage[caption=false]{subfig}
\usepackage{calc}
\usepackage{upgreek}

\usepackage{graphicx}% Include figure files
\usepackage{dcolumn}% Align table columns on the decimal point
\usepackage{bm}% bold math
\usepackage{paralist} % compact itemize
\DeclareMathOperator{\mc}{\enspace ,}

\DeclareMathOperator{\mf}{\enspace .}

\usepackage[colorlinks,citecolor=blue,linkcolor=blue]{hyperref}
\usepackage[]{cleveref}
\Crefname{figure}{Fig.}{Figs.}
\Crefname{table}{Tab.}{Tabs.}
\Crefname{equation}{Eq.}{Eqs.}
\begin{document}
\preprint{}

\title{Predicting Network Congestion by Extending Betweenness Centrality to Interacting Agents}

\author{Marco Cogoni}
\author{Giovanni Busonera}%
\affiliation{CRS4}%

\date{\today}

\begin{abstract}
    We present a simple model to predict network activity at the edge level, by extending a known approximation method to compute Betweenness Centrality (BC) with a repulsive mechanism to prevent unphysical densities. By taking into account the strong interaction effects often observed in real phenomena, we aim to obtain an improved measure of edge usage during rush hours as traffic congestion patterns emerge in urban networks. In this approach, the network is iteratively populated by agents following dynamically evolving fastest paths, that are progressively attracted towards uncongested parts of the network, as the global traffic volume increases. Following the transition of the network state from empty to saturated, we study the emergence of congestion and the progressive disruption of global connectivity due to a relatively small fraction of crowded edges.
    We assess the predictive power of our model by comparing the speed distribution against a large experimental dataset for the London area with remarkable results, which also translate into a qualitative similarity of the congestion maps. Also, percolation analysis confirms a quantitative agreement of the model with the real data for London. For seven other topologically different cities we performed simulations to obtain the Fisher critical exponent $\tau$ that showed no common functional dependence on the traffic level. The critical exponent $\gamma$, studied to assess the power-law decay of spatial correlation, was found inversely proportional to the number of vehicles both for real and simulated traffic.
    This simulation approach seems particularly fit to describe qualitative and quantitative properties of the network loading process, culminating in peak-hour congestion, by using only topological and geographical network features.
\end{abstract}

\maketitle

\section{Introduction}
\label{sec:intro}

One of the networking metrics that have been used the most in recent years to analyze network behavior is the
\textit{Betweenness Centrality} (BC). This measure of relative importance among the constituents of a network during a simultaneous peer-to-peer linkage process originated in the study of relations among people and abstract ideas~\cite{white_betweenness_1994}, but it has been applied with some success to transportation infrastructures such as airlines, cargo-ship, power grids and computer networks~\cite{barthelemy2011spatial,holme2003congestion,guimera2005worldwide,kazerani2009can}. It is roughly defined as the total flow passing over each node of a network when enumerating all Origin-Destination
(OD) pairs and connecting them via shortest paths. The definition of BC is easily extended to edges by measuring edge usage instead of nodes and we will refer to this variant throughout the paper.~\cite{freeman1991centrality}.
A major limitation of standard centrality-based approaches is that they can often only grasp a limited picture of a real network under stress. This is due to the fact that they rely on two main assumptions~\cite{agryzkov2019variant,gao2013understanding}:
(1) edge usage has fixed costs, and (2) unlimited agents can share the transport infrastructure regardless of physical capacity~\cite{cogoni2017ultrametricity}.
BC and other centrality measures have been used to predict which edges are subject to
the highest traffic demand, but the correlation with simulated or real traffic tends to vanish in the high-density regime~\cite{holme2003congestion} and in the presence of phenomena that are not explainable just with geographical or topological features~\cite{guimera2005worldwide}. Edge usage obtained from BC, in fact, typically mimics a low-density state of the network that usually happens with very small (compared to network geography) or very fast (with respect to congestion buildup timescales) agents~\cite{kazerani2009can}.

In this work, we aim to improve upon the standard BC by taking into account the strong interaction effects observed in real networks, in order to obtain a better measure of edge congestion. Our methods will be developed for urban transportation, using an approach that is easily transferable to other contexts~\cite{yeung2012competition, hamedmoghadam_percolation_2021,olmos_macroscopic_2018,colak_understanding_2016}.

Urban networks have been widely studied in recent years~\cite{colak_understanding_2016,li_percolation_2015,kirkley_betweenness_2018}, both in terms of their growth over time and of their complex dynamics under different traffic
levels. Network science has considerably helped to improve our understanding
of cities and to analyze and predict the reaction of
the different parts of the network under stress~\cite{hamedmoghadam_percolation_2021}. Such predictive analyses may be performed by using models depending just on the geographical and topological features of a city avoiding experimental traffic data, after careful validation against observations~\cite{bongiorno_vector-based_2021,lee_morphology_2017}.
Urban networks belong to the special class of (almost) planar graphs~\cite{diet_towards_2018,helbing_traffic_2001} whose topology is constrained by
the geographical embedding. This severely hinders their long-range connectivity
and also limits their maximum node degree~\cite{aldous2013true}. 
Since the study of node degree distributions alone cannot be expected to significantly improve our understanding of cities, non-local higher-order metrics such as network centralities
have been widely used both for theoretical studies and for practical
applications with notable success~\cite{white_betweenness_1994,newman_measure_2005}.

In this article (\cref{sec:methods}), we present a method where agents iteratively populate the network along dynamically evolving fastest paths. These paths are gradually pushed towards uncongested areas of the network as global traffic volume rises. We examine the transition of the network state from empty to saturated, investigating the emergence of congestion and the gradual disruption of global connectivity caused by a relatively small fraction of crowded edges.
Being still an open question how people plan their routes when driving in urban networks, our moving agents will be modeled as a mixed population of self-driving cars and human drivers using real-time traffic information. In this context, our results (\cref{sec:results}) that compare synthetic and real data show remarkable improvements with respect to standard BC: Whole network speed distributions agree quantitatively and spatial patterns of congestion are qualitatively very similar. The qualitative comparison is supported by analyzing how the spatial correlation of congestion decays with distance. Percolation analysis also produces compatible results for critical exponents within different traffic conditions.

\section{Methods}
\label{sec:methods}
In real urban networks, the average travel time is of the same order of magnitude as the typical timescale of congestion buildup. The strength of the interaction among vehicles depends on their local density and on the duration of the network loading process during rush hour: slower vehicles stay on edges longer, so their occurrence probability in a given road segment during the observation period is larger. Our approach stems directly from this fact and also takes into account that peak-hour periods are limited in time, usually lasting about one hour~\cite{li_percolation_2015}. This allows estimating the cumulative traffic seen on the roads during that finite time window. Thus, we propose a pseudodynamical model taking into account the contribution to traffic at a road-segment level due to each vehicle added to the network, by updating travel times at each step. The approach could, in principle, be extended to model the congestion decay that occurs when traffic volume eventually decreases, by removing vehicles reaching their destinations.

From the vast literature on transportation, we choose one of the simplest
models to describe vehicular behavior depending on the edge
physical properties and on the dynamical network state: the single regime
Greenshields model~\cite{helbing_traffic_2001,rakha2002comparison},
for which speed starts as free flow on an empty road, decreasing linearly
to zero with maximum density. To complement the traffic model, we impose a selfish behavior on our vehicles: they follow the fastest path (not the shortest one) as computed at the time of leaving their origin node.

\subsection*{Interaction Model}
We simulate the network evolution, as observed by travelers, while the traffic increases from zero up to almost complete gridlock, signaled by the vanishing probability of adding new paths not containing congested edges. The traffic network is modeled as a directed, weighted graph whose edges, identified by $e$, represent road segments between adjacent intersections (nodes) and possess three constant features: physical length $l_e$, maximum speed $v^*_e$ and number of lanes $c_e$. Nodes are featureless.

The network traffic grows incrementally by activating one new path $\pi^i$ at each simulation step $i$, in order to reach the desired target value at time $T$ (the end of the simulation).
Thus, $i$ can be interpreted both as the current number of added paths and as a temporal marker within the sequence of OD pairs randomly generated for each simulation. This procedure mimics the well-known method of approximating BC~\cite{riondato2016fast}.
For simplicity, and to be able to compare results with respect to the standard BC, traffic will be added uniformly to the network. It is however straightforward to adapt our procedure to any OD matrix.

The state of the network at each timestep is defined by 
the temporal occupancy factors due to all vehicles added so far to each edge: $s_e^i = \sum_{j=1}^i \sigma_e^j$. The single-vehicle occupancy $\sigma_e^i$ is defined as the ratio between the time (crudely approximated) spent on edge $e$ and $T$:
\begin{align}
  \sigma_e^i &= \min\left(\frac{T_e^i}{T}, 1\right) \mc
  & \text{where } T_e^i &= \frac{l_e}{v_e^i} \mc
                           \label{omega_def}
\end{align}
with $v_e^i$ following a Greenshields linear law~\cite{jin2021introduction}:
\begin{equation}
  v_e^i = v_e^*(1-\rho_e^{i-1}) \mc
  \label{greenshields_def}
\end{equation}
where  
\begin{equation}
      \rho_e^{i-1} = \frac{s_e^{i-1} L}{l_e c_e} \in [0, 1] \mc
    \label{density_def}
\end{equation}
is the normalized vehicle density at the previous step, monotonically increasing with $i$. $L$ is the average space occupied by one vehicle, and the denominator represents the edge capacity. 
A non-interacting system, obtained in the limit $L\rightarrow 0$, approaches the state $s_e$ as computed with the standard BC.
The approximate time to travel along $\pi^i$ will be the sum over the edges $T_{\pi^i}=\sum_{e\in\pi^i}T_e^i$.
The total occupancy due to a single vehicle is $\sum_{e\in\pi^i} \sigma_e^i \le 1$ and will only reach $1$ for a path $\pi^i$ with total traveling time $T_\pi\ge T$.

In more detail, the state of the network is iteratively obtained ($s_e^i=f(s_e^{i-1})$), starting with unused edges ($s_e^0=0$ and $\displaystyle T_e^1 = \frac{l_e}{v^*_e}$) and according to the following dynamic process:
\begin{itemize}
\item a pair of OD nodes is chosen, independently and uniformly at
  random, and the fastest path $\pi^i$ connecting the nodes is computed, given the current travel times $T_e^i$;
\item starting from O, we assign the respective shares of occupancy $\sigma^i$ induced by $\pi^i$ during $T$, to each edge $e\in\pi^i$:
  \begin{equation}
    s_e^i = s_e^{i-1} + \sigma_e^i \mf
    \label{occup_update}
  \end{equation}
  Note that, as soon as the sum along $\pi^i$ of the added $T_e^i/T$ factors
  reaches $1$ (travel time longer than the simulation), we skip the remaining edges until the destination, to avoid
  increasing by more than a unit the vehicle occupancy along $\pi^i$;
\item $v_e^{i+1}$ and $T_e^{i+1}$ are updated according to
  \Cref{omega_def,greenshields_def}, using the new
  $\rho_e^i$ value;
\item this process is iterated until the target total traffic is reached.
\end{itemize}

Intuitively, the occupancy factor induced by a vehicle over an edge
$e$ is proportional to the time the vehicle is supposed to spend on it (${T_e}/{T}$), as forecast at its departure,
and the sum over the whole path will be equal to unity (certainty of
finding the vehicle within $\pi$ during $T$) only when
$T_\pi\ge T$.
Since the initial vehicles find a nearly empty network, their fastest paths
and travel times are virtually equivalent to the non-interacting case. With rising traffic, however, edges fill up and the previous fastest paths will
disappear and less-used roads and residential neighborhoods will be chosen.
Some edges will eventually reach maximum density and become congested.
If the fastest route from O to D comprises a congested edge
(i.e., the network is disconnected) we still choose to add the initial part of the path, but skip all remaining edges from the first congested one.
This allows us to model the backward propagation of traffic jams
observed at high traffic volumes~\cite{olmos_macroscopic_2018,taillanter2021empirical}.
The order in which OD pairs are selected for adding vehicles to the network can lead to different states, thus, multiple replicas of the system are simulated to verify the stability of the results. Replicas differ in the subset of OD pairs chosen, but no relevant differences in the average results were detected just by reshuffling the same subset of ODs.

\subsection*{Spatial correlation of congestion}
To estimate the degree of spatial correlation of congestion between edges, we adapt the definition of correlation used in modeling collective behaviors ~\cite{witten1981diffusion,cavagna2010scale}. We define:
\begin{equation}
    \label{eq:spatial_corr}
        C(t)=\frac{1}{c_0}\frac{\sum_{ij}\langle\rho^*_i\rho^*_j\rangle\delta(t-t_{ij})}{\sum_{ij}\delta(t-t_{ij})} \mc
\end{equation}

in which $\rho^*$ can assume the values $0$ or $1$ whether below or above the congestion threshold, respectively; $t_{ij}$ represents the fastest travel time between the two edges on the empty network; $\delta$ is a rectangular window function selecting times close to $t_{ij}$; and $c_0$ is a normalization factor. Angle brackets $\langle\cdot\rangle$ refer to the ensemble average over replicas, where replicas may be independent simulations with different OD pairs or multiple instances of real traffic states for the same time slot and weekday.
In our analysis, we will disregard the normalization factor $c_0$ since our focus is on the $\gamma$ exponent. We compute our correlations as a function of travel time rather than distance. Travel time on the empty network, in this context, is the preferred way to estimate proximity between edges, because Euclidean distance is known to lead to distortions, as it disregards connectivity. Moreover, we perform correlation analysis on raw speed values and not on their fluctuations~\cite{taillanter2021empirical}, as opposed to some previous studies~\cite{cavagna2010scale,petri2013entangled}. 

\section{Results and discussions}
\label{sec:results}
In order to validate our approach, we first compare the traffic properties of a large-scale real-world dataset with those obtained from our interacting network model. We show that, just by imposing a simplified repulsion mechanism among vehicle paths, the agreement with real data markedly improves with respect to the standard BC. The comparison is first performed between speed distributions over the whole network. Then, by correlating the speeds edge-to-edge, we identify the simulated traffic volume ($V$) best matching the real data at peak hours. Since speed depends on road usage (vehicle density), we also compare the congestion map obtained from BC and from our model against measured data, illustrating how the proposed method seems to be able to reproduce realistic congestion states for high-volume traffic situations. We also compute the spatial correlation for edge congestion to verify whether the $\gamma$ critical exponent is comparable to previous results for different traffic levels. Since another property characterizing the network is the cluster-size distribution during critical percolation, which is known to follow a power law with exponent $\tau$, we also study its value for increasing traffic volumes and compare it against the real traffic dataset at different hours.

\paragraph*{Comparison with real-world measurements --}
To understand how realistic the results produced by the present model are, we compare them against a large experimental dataset provided by UBER~\cite{UberMovementData}. The dataset provides GPS tracks for taxi fleets in several metropolitan areas, but only the data for the city of London was used in this work, discarding the others for insufficient sampling.
All road networks were obtained from OpenStreetMap (OSM) by using the OSMnx library and downloaded in their latest state, except in the case of London for which we matched the period of the recorded real traffic dataset. The UBER Movement dataset contains data recorded for a high fraction of the street segments at hourly intervals for several years: we selected the first six months of 2019 to avoid spurious effects due to the pandemic. Speed values in the dataset are always one-hour averages from multiple vehicles since single-vehicle data are not available due to privacy concerns. For our analyses, we define a speed factor for every edge in the network as the ratio between the hourly average speed and the maximum value ever observed in that edge during the whole period.
Since speed data can be sparse especially during low traffic (see Supplemental Material (SM) - Fig.~S1a) and roads typically consist of several segments, we also derived a spatially coarser dataset with fewer missing values. This was done by averaging speed on multiple edges belonging to a single street at the same time. The use of this coarser dataset will be specified in the results. Throughout the paper, we will focus on four specific times of the workday: peak hours, 8-9 and 17-18, and off-peak, 10-11 and 22-23. See Fig.~S1b for the slowdown behavior during the $180$ days of the dataset and how workdays differ from the weekend, and how even Sundays are distinguishable from Saturdays.
\begin{figure}[htb]
    \includegraphics[width=\linewidth]{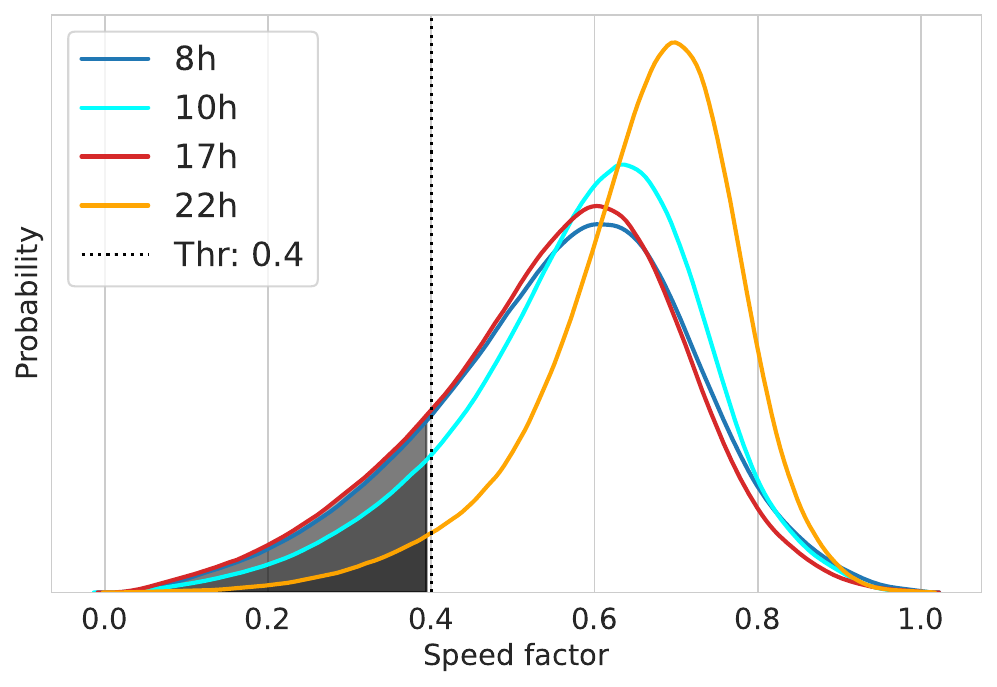}
    \caption{\label{fig:speed_distrib}UBER speed factor distributions. Shaded areas on the left highlight congested edges.}
\end{figure}
In \Cref{fig:speed_distrib}, we compare the speed factor distributions over the whole network for the selected time slots: the orange curve is associated with the nocturnal time slot (22h) and, as expected, it has the highest average speed and the smallest amount of edges with slowdowns below $40\%$ of the free flow. We choose this value as a threshold to define congested edges. The morning off-peak distribution (10h, cyan) shows a considerably slower average speed factor than the evening one and it is globally more similar to the 8h (blue) and 17h (red) that are almost indistinguishable in this respect.

\paragraph*{Urban network selection and simulation details --}
To complement the comparisons with real data, we select the vehicular transportation layer of the urban networks of eight large cities (five in Europe, two in the US, and one in China) and their surroundings from OpenStreetMap. The cities have been selected to be representative of very different urban structures, stemming from their different history, geographic location, and local site features. The radius of the circle inscribed in each square region is $20$ km for all cities except for Rome and Madrid ($12$ and $15$ km, respectively). The number of edges of the corresponding graphs goes from about $1.0\times10^5$ for Rome to about $5.5\times10^5$ for London. Detailed information concerning OSM road networks is reported in Tab. S1 of the SM. The final number of added OD pairs at the end of each simulation is $V = 2.0\times10^6$, sufficient to bring all cities to a deeply congested state, as shown in \Cref{fig:incomplete_path}. The computed vehicular speed on each edge is multiplied by a small Gaussian noise $\xi(\mu=1,\upsigma=0.1$) to reproduce the intrinsic variability of drivers. All simulations were run with $T=3600$~s.

\paragraph*{Percolation analysis --}
Being able to produce synthetic traffic with adjustable congestion levels, we perform a critical percolation analysis to check open questions about the fragmentation process under stress: the network graph is pruned at increasing speed thresholds, to locate the phase transition exactly when the size distribution of the resulting strongly connected subnetworks follows a power law~\cite{cogoni2021stability}. The critical exponent associated with this transition is supposed to depend, in particular conditions, on traffic intensity, as observed for real-world datasets in large cities such as Beijing~\cite{zeng_switch_2019}. It is also supposed to show metastability during rush hours~\cite{zeng_multiple_2020}.
The above percolation phase transition is not a property of high congestion alone: it is an indicator of a change in the network behavior visible at all traffic levels, but appearing for different speed thresholds. Within this percolation paradigm, even some of the free-flowing edges of an almost empty network would be classified as dysfunctional at criticality~\cite{li_percolation_2015,zeng_switch_2019,cogoni2021stability}. On the other hand, when dealing with real traffic congestion, edges are customarily deemed dysfunctional only with road densities approaching their physical limit and diverging travel times~\cite{hamedmoghadam_percolation_2021}.

\paragraph*{Model approximations --}
The proposed method has been developed to evaluate the effects of adding agent interaction on the network while keeping the model as simple as possible. The known approximations are mainly due to: (1) the Greenshields model is very rough, especially for urban traffic, but its simplicity helps the explainability of the results; (2) no explicit time evolution as in cellular automata models (e.g., Nagel-Schreckenberg), but a sequence of path additions in a \emph{first come, better served} approach; (3) OD pairs are uniformly extracted, thus preventing us to fully describe the known switch of the average traffic direction from the morning to the evening peak hours~\cite{colak_understanding_2016,mazzoli2019field}.

\subsection*{Comparing the interacting model, BC, and real traffic}
\begin{figure}[htb]
    \includegraphics[width=\linewidth]{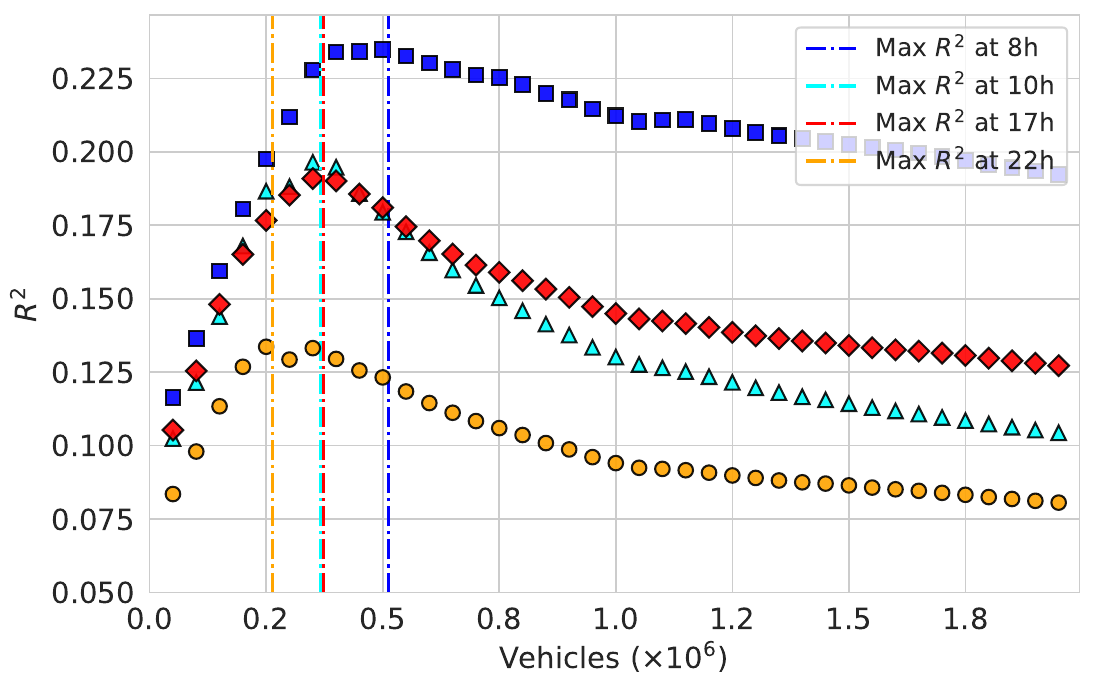}
    \caption{\label{fig:uber_sim_corr}Edge-level correlation between real speed data (8h, 10h, 17h, and 22h) and simulated speed for increasing traffic volume for London.}
\end{figure}

We first consider the correlation between the speeds observed in coarse-grained real traffic during the four time slots and those produced by the simulation for increasing volumes of traffic. We use Ordinary Least Squares to compute the coefficient of determination $R^2$ (square of the Pearson correlation coefficient), which quantifies the ratio of explained variance.
In \Cref{fig:uber_sim_corr}, each correlation curve shows a clear maximum: at 8h (blue), $R^2\sim0.23$ at about $V = 0.5\times10^6$ vehicles after a steep growth from lower traffic. After reaching the maximum, the correlation slowly decays to $R^2\sim0.2$ for very high congestion. For different time slots, the maximum correlation is lower (e.g., $R^2\sim0.2$ at 10h (cyan) and 17h (red), while $R^2\sim0.13$ at 22h (orange)) and happens at lower traffic volumes. During off-peak time slots $R^2$ (red and orange) also decays much more for high simulated traffic. These values should be compared with the highest $R^2$ obtained for BC and real traffic, which is of the order of $3\times10^{-2}$.

\begin{figure}[htb]
    \includegraphics[width=\linewidth]{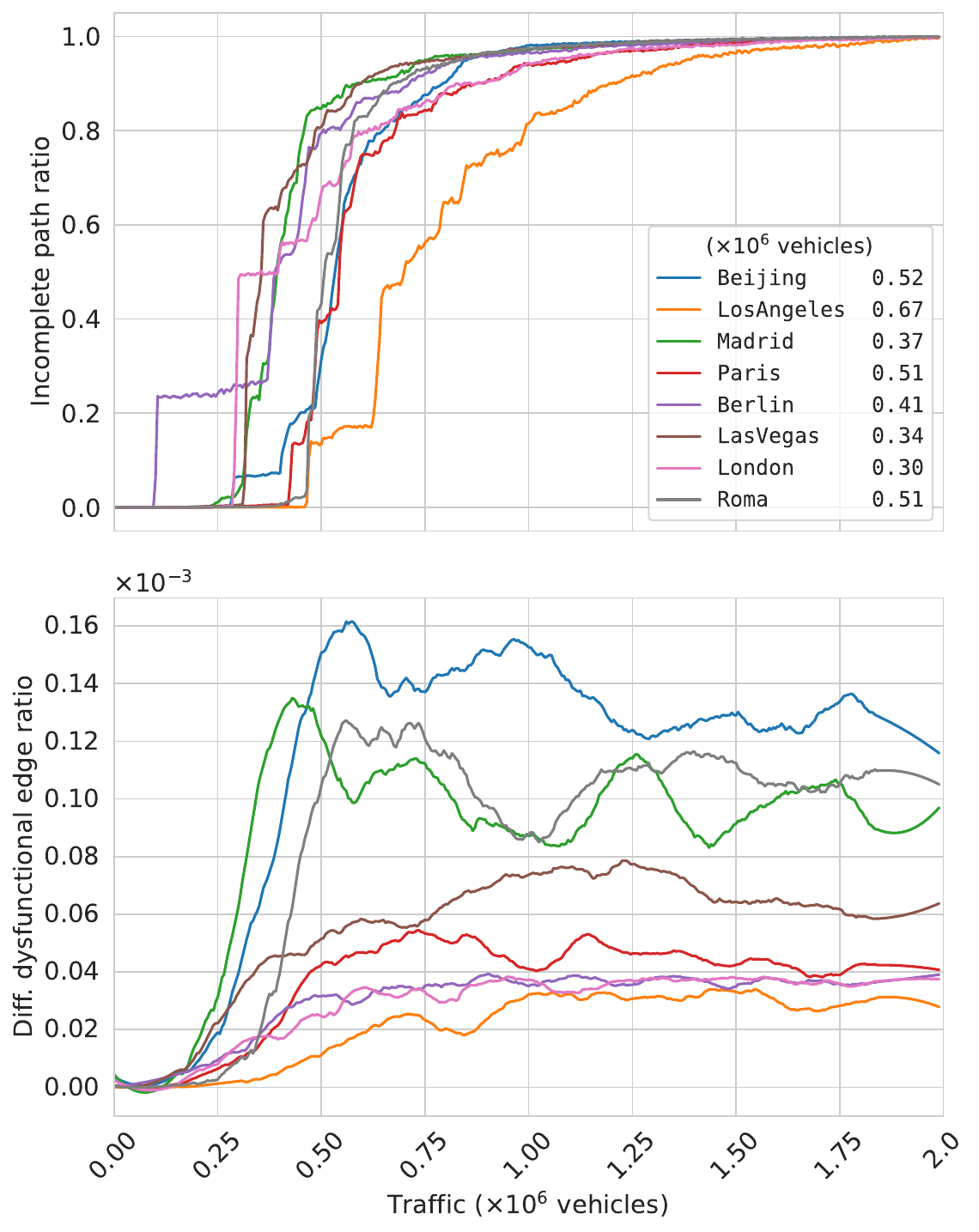}
    \caption{\label{fig:incomplete_path}Simulation: on the top, the fraction of incomplete paths follows a sigmoid-like curve as traffic load increases. $V^*$ is shown in the legend for all cities. Each curve is averaged over five replicas. At the bottom is shown the fraction of edges removed at each step from city networks. The area under each curve is the total fraction of dysfunctional edges.}
\end{figure}

\Cref{fig:incomplete_path}(top) shows how each urban network reacts to increasing levels of traffic: most cities behave in a similar way with respect to the fraction of ``impossible'' paths -- the ones between OD pairs that can be connected only including congested edges -- with a sigmoid-like curve that visualizes the progressive network breakup. We refer to the number of vehicles associated with the sigmoid center as the critical traffic volume $V^*$.
Los Angeles (orange curve) appears to be more resistant and its network breaks after adding about $0.7\times10^6$ vehicles, a fact due in part to the total length of its roads and to the mesh-like topology that produces a strong path degeneration (multiple options at a similar cost). Las Vegas (brown), on the other hand, albeit sharing a similar organization, is much smaller and collapses together with the rest of the European cities. Berlin, London, Las Vegas, and Madrid are the first to fail (traffic volume $V \sim0.35\times10^6$) followed by Beijing, Rome, and Paris, resisting up to $V \sim0.5\times10^6$ despite a vastly different total street length within the two groups.
%% SIMULATED INCOMPLETE PATHS for all cities

The traffic volume needed to reach maximum $R^2$ is just above the value necessary to split into two halves the London network in the simulation, as visible in \Cref{fig:incomplete_path}: the pink curve has a small plateau for about $V = 0.3\times10^6$ vehicles when all bridges on the River Thames become congested, making it impossible to reach the other side of the city within the chosen $T$. 
\Cref{fig:incomplete_path}(bottom) shows the fraction of edges that become dysfunctional at each step and highlights the fact that a very small minority of congested edges can lead to transportation breakdown: Beijing, Rome, and Madrid remain connected with a total fraction (area under each curve) of defects much larger than the other cities. Notably, the two American cities with mesh-like geometry collapse with relatively fewer congested edges. 
In Figs.~S22-S23 we also report detailed graphs for all cities containing the incomplete path ratio, its flex location, and the curve of the fraction of removed edges at each step.

%% Edge-level speed comparison UBER vs sim
\begin{figure}[htb]
    \includegraphics[trim={0 0 0 0cm},clip,width=\linewidth]{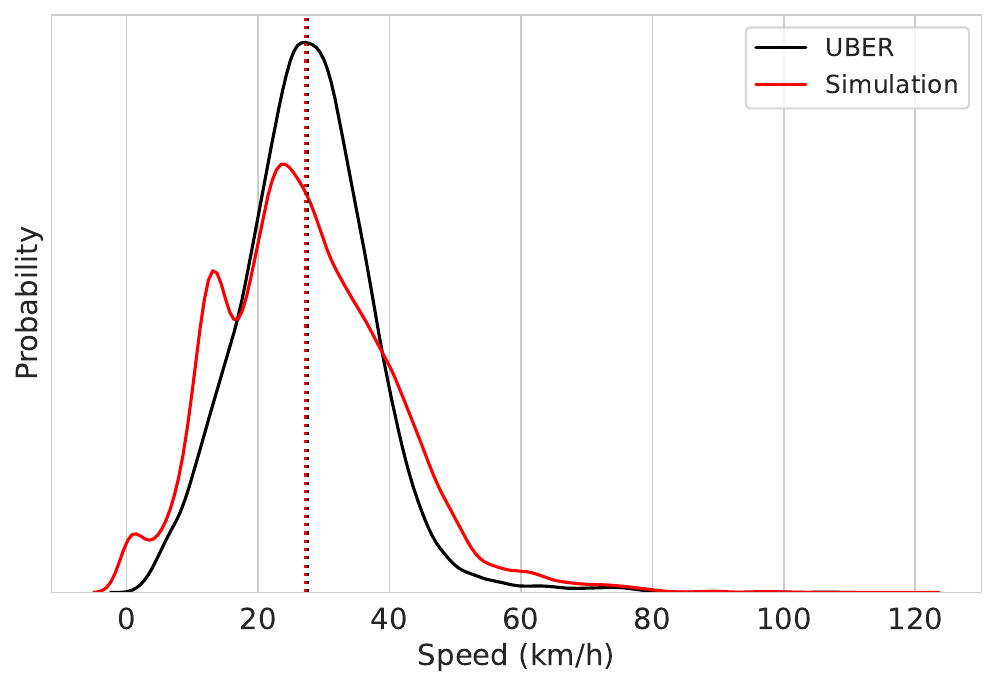}
    \caption{\label{fig:uber_vs_sim_speed_fact_distr}Speed factor distribution comparison for the real dataset (at 8h) and the simulation (at maximum $R^2$).}
\end{figure}
%% Uber Images
\begin{figure}[htb]
    \includegraphics[width=\linewidth]{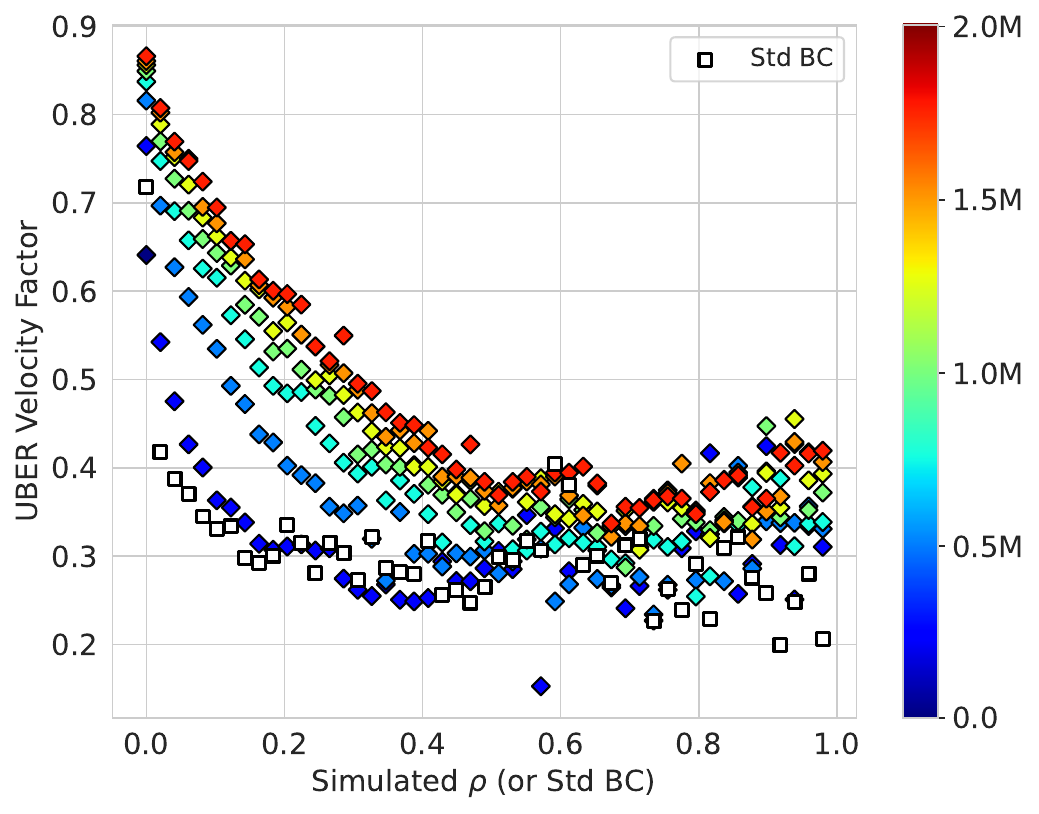}
    \caption{\label{fig:uber_speed_fact}Real dataset (at 8h) speed factor dependence on simulated $\rho$ for different levels of traffic load (identified by the color). Each symbol is the average real speed factor for all edges within a $\rho$ bin. Simulated densities show a correlation for values up to a density of 0.5 whereas the BC is totally uncorrelated after 0.25.
    For reference, the empty black squares show the normalized standard BC prediction.}
\end{figure}

In \Cref{fig:uber_vs_sim_speed_fact_distr} the whole-network speed distribution obtained from our model (at maximum $R^2$) is compared to the real dataset during the morning peak hour: it shows a remarkable overlap between the two curves and almost identical average values. The small superimposed peaks are remnants of the original speed limits on uncongested roads.
\Cref{fig:uber_speed_fact} presents a set of curves showing how the real data (at 8h), for all edges (speed factor, $y$-axis), is connected to the simulated density ($\rho$, $x$-axis) at different levels of synthetic traffic: dark blue symbols correspond to $V \sim0.2\times 10^6$ vehicles, light blue to $V \sim0.5\times 10^6$, cyan $V \sim0.7\times 10^6$, and red to $V \sim2.0\times 10^6$. They are a subset of all points shown for the blue curve in \Cref{fig:uber_sim_corr}. Empty black squares show how poorly the normalized standard BC correlates with the real speed data and how our model with very low traffic (dark blue curve) converges to BC\footnote{The $x$-axis values refer to the center of bins of either density $\rho$ or standardized (std) BC: given a specific $x$ value, all edges with $\rho$ (or std BC) within the interval $[x-\delta,x+\delta]$ are grouped, then their corresponding UBER speeds are averaged and the mean value is plotted on the diagram.}.

After comparing speed distributions and computing their correlation levels, we turn to a qualitative comparison between traffic maps, where each street edge is characterized by a normalized quantity that can be considered a proxy for congestion: (1) load computed by standard BC, (2) speed factor from the real dataset, and (3) vehicular density from the simulations.
\begin{figure*}[htb]
    \begin{minipage}[b]{\textwidth}
        \subfloat[\label{fig:london_maps_a}]{%        % FULL MAP
            \includegraphics[trim={10cm 10cm 10cm 10cm},clip,width=0.31\linewidth]{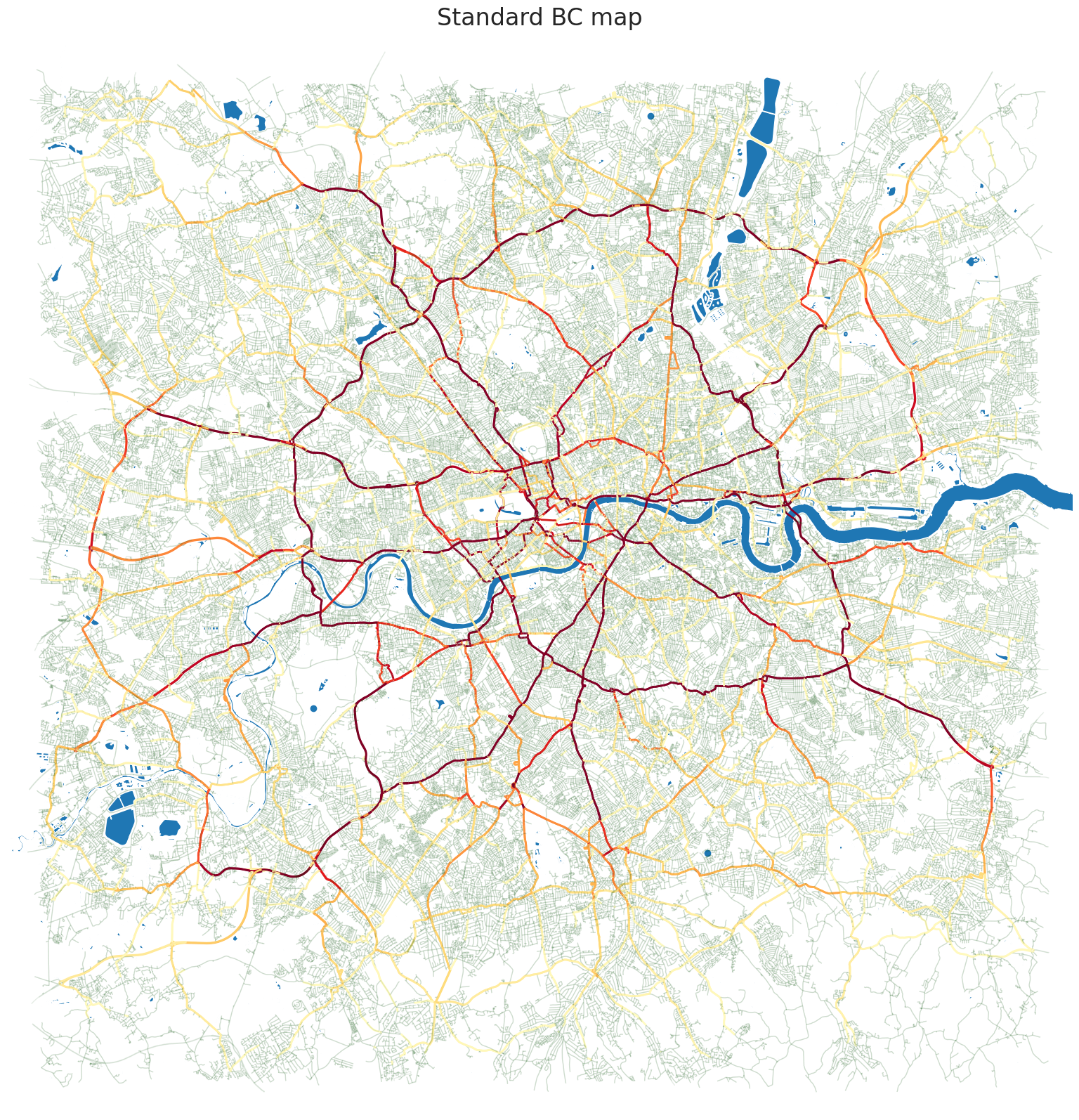}
        }\hfill
        \subfloat[\label{fig:london_maps_b}]{%
            \includegraphics[trim={10cm 10cm 10cm 10cm},clip,width=0.31\linewidth]{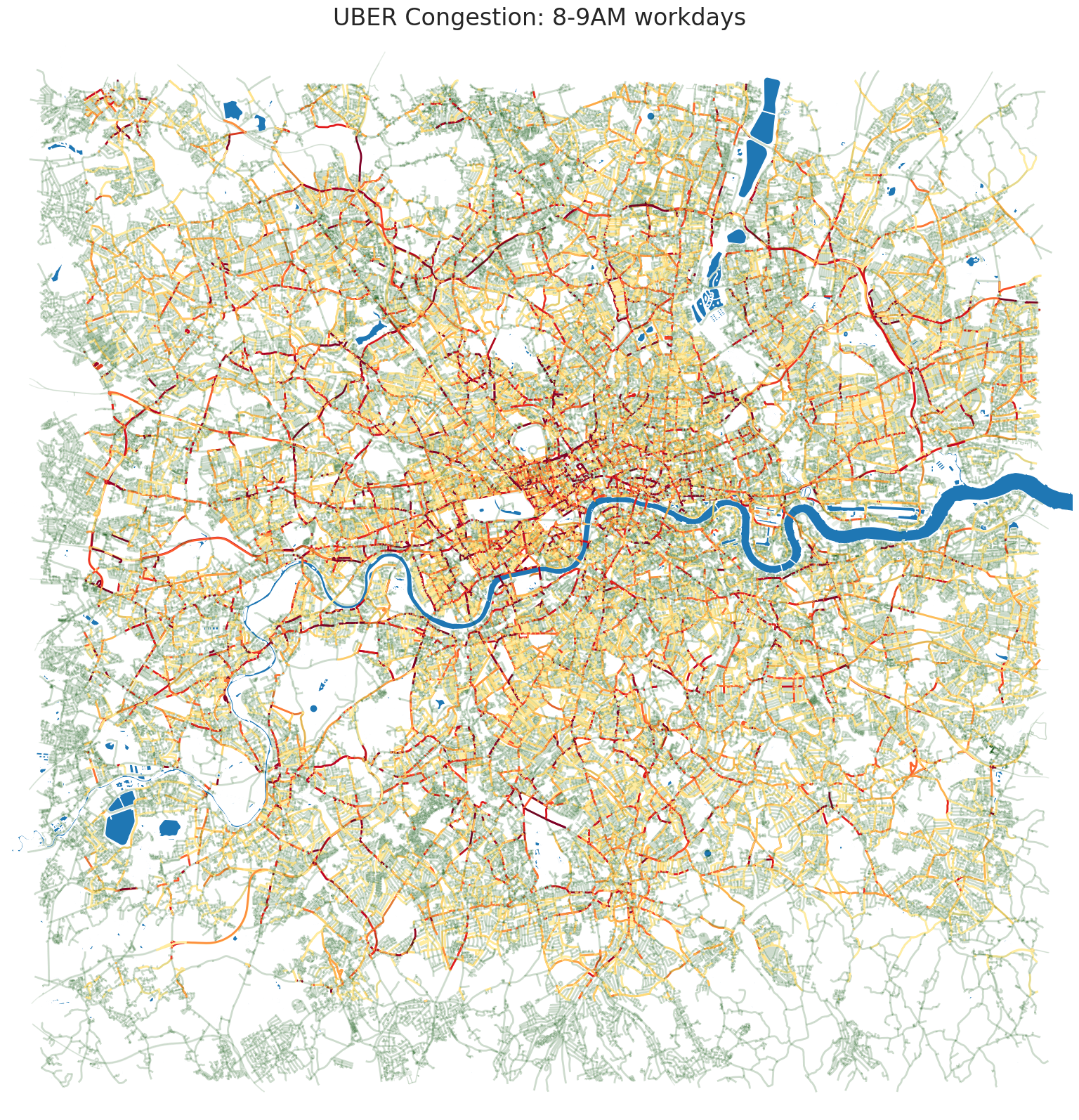}
        }\hfill
        \subfloat[\label{fig:london_maps_c}]{
            \includegraphics[trim={10cm 10cm 10cm 10cm},clip,width=0.31\linewidth]{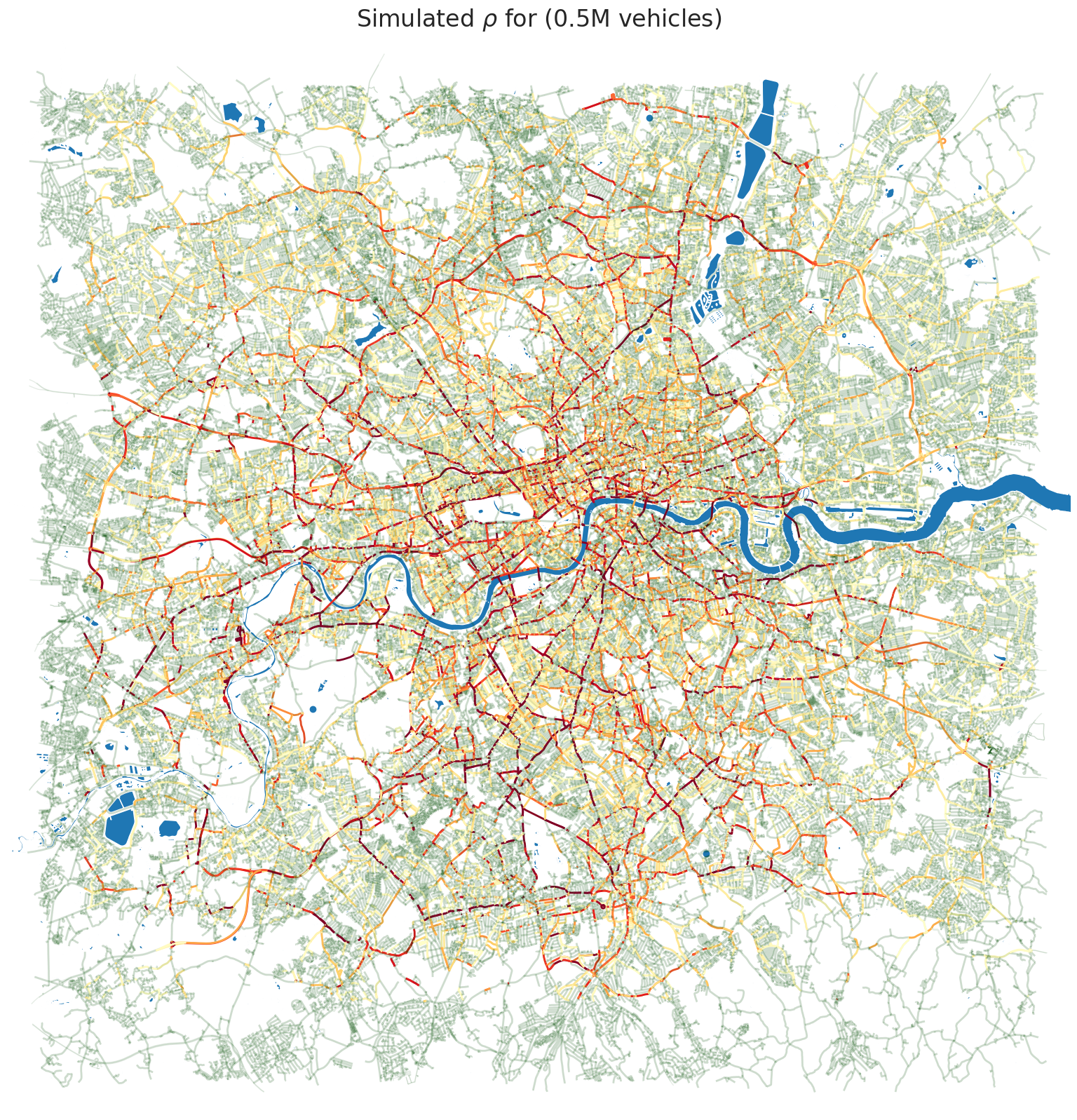}
        }
        \caption{\label{fig:london_maps}Central London congestion maps: (a) normalized standard BC, (b) real dataset inverse speed factor at 8h, and (c) simulation traffic density ($\rho$) at maximum $R^2$ (with $V = 0.5\times 10^6$). Yellow is low road usage and red stands for the $95$-th percentile for each distribution.}
    \end{minipage}
\end{figure*}
In \Cref{fig:london_maps} we compare, side-by-side, the standard BC, the UBER traffic speed factor during the morning rush hour, and the vehicular density ($\rho$) of simulated traffic for the central part of London. The standard BC (\Cref{fig:london_maps_a}) badly underestimates congestion on most edges with respect to the real network state as shown in \Cref{fig:london_maps_b}, except for some important arteries that are correctly identified. This behavior is expected since BC does not take into account the interactions and, therefore, it is not able to describe the progressive traffic spillover towards secondary (but still functional) roads. 
\Cref{fig:london_maps_c} shows that, in our simulations, the traffic volume that correlates the most with the real data at 8h is $V = 0.5\times 10^6$, which is $\sim70\%$ above $V^*$ (see \Cref{fig:incomplete_path}).
Thus, for the morning peak hour, our model (\Cref{fig:london_maps_c}), qualitatively and quantitatively, far outperforms the BC at describing real traffic patterns. The prediction quality breaks down towards the edges of the simulated area because no transport is simulated outside of it (full map shown in Fig.~S10).
We compare the congestion maps obtained from the simulations for all other cities, for $V^*$ and at half volume $\frac{1}{2}V^*$, along with the BC prediction. The simulation at $\frac{1}{2}V^*$ shows that in a low interaction scenario, the BC qualitatively resembles the results of our model, while at higher traffic levels its pattern is only able to highlight the main roads. It is especially enlightening to observe the results for Los Angeles and Las Vegas which are characterized by a mesh-like topology with a high level of shortest path degeneracy: for low congestion a few roads attract most of the traffic, which then spills over to the alternative routes with equivalent lengths for higher $V$. The existence of these alternative paths guarantee a higher network resilience but at the cost of exploiting residential areas, which have been reported to be already experiencing a growth in congestion and noise as the use of traffic-aware automatic route planners becomes widespread among drivers~\cite{batac2022shortest}.
Older European cities behave differently since topology is much more complex and stratified, and residential areas are more protected than their US counterparts by the existence of a deeper road hierarchy, able to sustain the effort during peak hours. All these results are shown in Figs.~S2-S9.

\begin{figure}[htb]
    \includegraphics[trim={0 0 0 0cm},clip,width=0.89\linewidth]{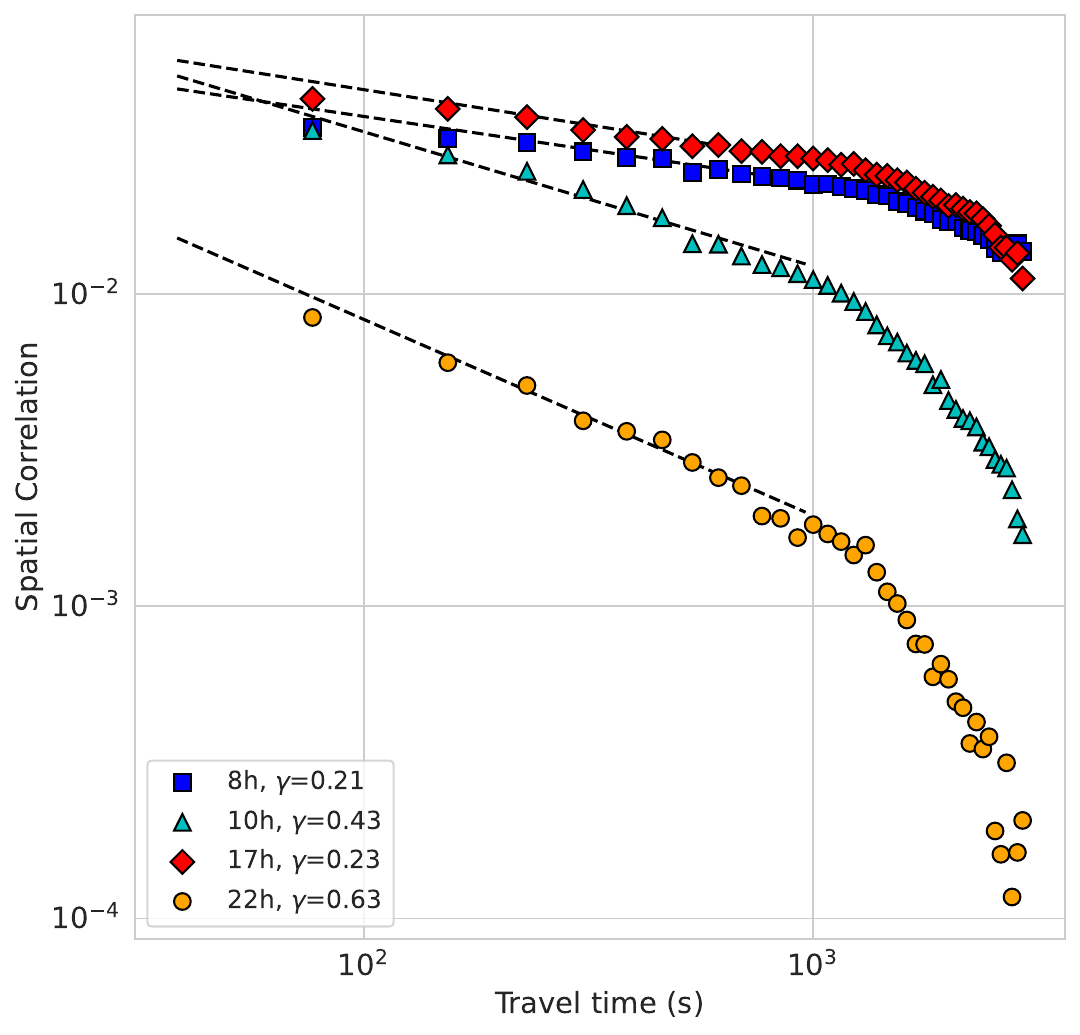}
    \caption{\label{fig:uber_connected}London spatial correlation of congestion for all four time slots using the UBER data. Regressions are performed on the linear part of the curve and their $\gamma$ exponents are shown in the legend for each time slot.}
\end{figure}

\begin{figure}[htb]
    \includegraphics[width=\linewidth]{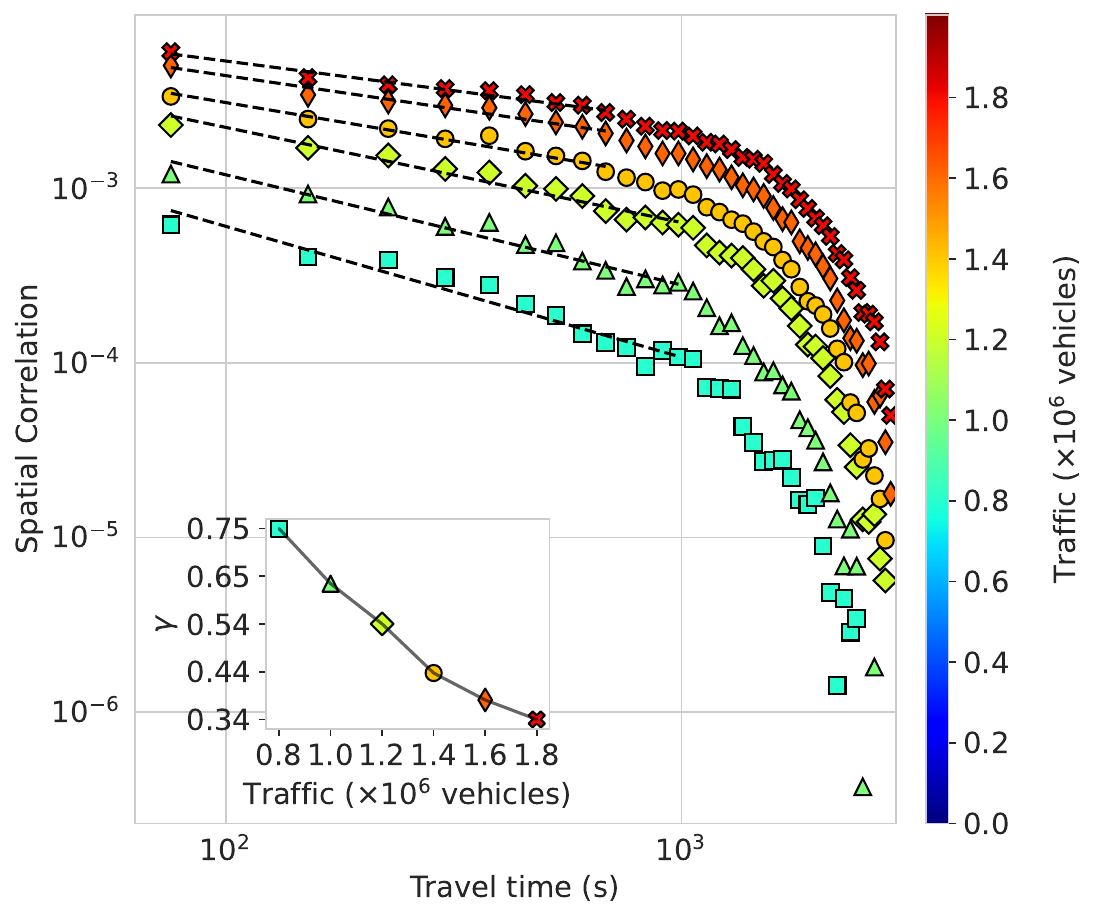}
    \caption{\label{fig:sim_connected}Simulation (London) of the spatial correlation of congestion for increasing traffic load. Linear regressions and the associated $\gamma$ exponents are computed for the linear part of the curve and $\gamma$ dependence on traffic level is shown in the inset.}
\end{figure}

We now turn our attention to the spatial correlation of congestion for both simulated traffic and real data. In \Cref{fig:uber_connected}, we show the spatial correlation computed by applying \Cref{eq:spatial_corr} on congestion maps such as those depicted in \Cref{fig:london_maps_b} for UBER data at four-time slots, two peak-hour and two off-peak. It is clear that the correlation follows a power law until the exponential tail discussed in Ref.~\onlinecite{taillanter2021empirical} kicks in for long distances. An interesting result is that the relevant exponent has two regimes: $\gamma\sim0.2$ for peak hours and $\gamma\sim0.5$ for off-peak periods. This means that congestion decreases much faster with distance (as usual, measured in travel time on an empty network) for lighter than for heavier traffic. This result goes partially against previous findings that detected no difference in the $\gamma$ values for different traffic levels~\cite{taillanter2021empirical, petri2013entangled, daqing2014spatial}.
Performing the same analysis on the synthetic data shows a trend similar to the real dataset as shown in \Cref{fig:sim_connected}: steep slopes are associated with low and medium traffic ($\gamma\sim0.8$ for $V \sim0.5\times 10^6$) and flatter profiles to highly congested networks ($\gamma\sim0.4$ for $V \sim1.5\times 10^6$). So the simulation is compatible with the real data for $\gamma$ in similar traffic scenarios.

The same analysis was performed for all the other cities, for which the spatial correlations show similar behavior to what was observed for London: in most cases, $\gamma$ decreases linearly with growing traffic except for Beijing, for which we see an abrupt regime switch: from $\gamma\sim0.4$ for $V \sim0.6\times 10^6$ to a constant $\gamma\sim0.2$ for higher $V$. More details about the spatial correlations and the associated co-congestion probabilities are presented in Figs.~S17-S20 and Fig.~S21. 

\begin{figure}[htb]
    \includegraphics[width=\linewidth]{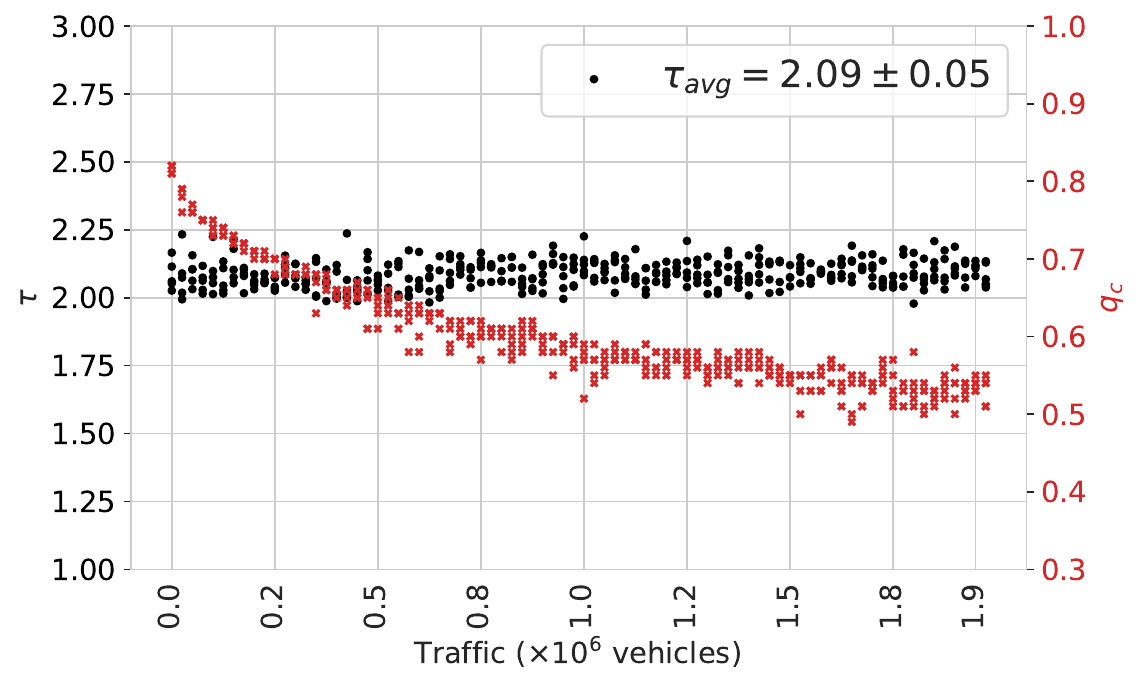}
    \caption{\label{fig:tau_qc}Simulation (London) of the critical speed factor $q_c$ (red) and critical exponent $\tau$ (black) for increasing levels of traffic load. Multiple points for the same abscissa represent different replicas. $\tau_{avg}$ is the average over the whole domain for all replicas.}
\end{figure}

The last comparison between real and simulated traffic regards percolation results. Details on the percolation methods are given in the SM. For each traffic level, both real and synthetic, we first compute the critical speed factor $q_c$ at which the size distribution of the graph-connected components becomes a power law. Once $q_c$ is known, a linear regression produces the $\tau$ value. In \Cref{fig:tau_qc} we see that $\tau$ stays almost constant over the whole range of simulated traffic volumes and its average value is $\tau_{avg}=2.09\pm0.05$, a result that agrees with the theoretical value for isotropically directed graphs of $\tau\sim2.1$~\cite{noronha2016perc_directed_isotr_lattice}. 
This result is slightly above the value obtained for real traffic data: $\tau$ varies very little for extremely different congestion states and stays just below $2.0$ as shown in \Cref{tab:uber_tau_qc}. The critical threshold speed factor $q_c$ for the simulations, on the other hand, decreases from about $q_c\sim0.8$ for no congestion to $q_c\sim0.55$ for a saturated network. These values approximate well what we observe for real traffic: $q_c\sim0.58$ ($0.66$) for off-peak, and $q_c\sim0.54$ ($0.51$) for peak hours, for morning (afternoon) time slots.
%% London UBER percolation results
\begin{table}
        \begin{tabular}[t]{c@{\hskip .1in}c@{\hskip 0.1in}c@{\hskip 0.1in}c}
            \hline
            time & \ $q_c$ & $\tau$ & $\gamma$\\ 
            \hline
            \noalign{\vskip 1mm}
            $8-9$ & $0.54\pm0.05$ & $1.95\pm0.03$   & $0.21\pm0.05$\\
            $10-11$ & $0.58\pm0.02$ & $1.99\pm0.03$ & $0.43\pm0.05$\\
            $17-18$ & $0.51\pm0.02$ & $1.96\pm0.03$ & $0.23\pm0.05$\\
            $22-23$ & $0.66\pm0.01$ & $1.96\pm0.02$ & $0.63\pm0.05$\\ 
            \hline
        \end{tabular}
\caption{\label{tab:uber_tau_qc}Critical speeds and critical exponents for cluster size distribution and for spatial correlation measured at four time slots for the UBER dataset for London.}
\end{table}

The main result here is that we observe no clear $\tau$ dependence on traffic volume for London, both for simulations and real traffic, moreover, their $q_c$ trend shows a good agreement. 
Most of the other cities have very small $\tau$ dependence on traffic volume and its average value agrees with the theoretical prediction of $\tau\sim2.1$ with two notable exceptions: (1) Los Angeles starts from $\tau\sim2.15$ for low traffic, but rises to $\tau\sim2.2$ for high $V$; (2) Beijing, on the other hand, has a baseline $\tau\sim2.0$, but has at least one clear peak occurring for low-medium traffic volumes with $\tau\sim2.3$ for all replicas. For Beijing, $q_c$ confirms this anomaly with a clear dip at the same traffic volume. This could be related to the previous results of Ref.~\onlinecite{zeng_switch_2019} in which a switch for $\tau$ was observed between peak hours and off-peak time slots with values similar to ours. This result needs further analysis, but it would be very interesting if such a dynamical effect could be predicted just by starting from city maps. Berlin showed another interesting behavior: $\tau$ starts with a relatively high value of $\sim2.1$ up to $V\sim0.5\times10^6$, where it clearly drops to $\tau\sim2.0$, then stabilizing to $\sim2.05$ afterward. This dip in $\tau$ is also visible for $q_c$ and is associated with a strong increase in replica variability. See Figs.~S11-S14 for the detailed graphs of all cities. 

\section{Conclusion}
\label{sec:conclusion}
In this work, we introduced a simple model to predict network activity at the edge level. Our method is inspired by a known and intuitive method to approximate BC, but also introduces a repulsive term preventing unphysical densities. This approach leads to dynamically evolving fastest paths, that are progressively attracted towards uncongested parts of the network, as the total traffic volume increases. This simulation scheme is particularly fit to describe the network loading phase leading to peak congestion. 

We extensively compared our predictions for the Greater London area to massive measured data of real traffic speeds, finding a notable agreement in particular for speed distributions over the whole network at specific time slots. The qualitative accord for congestion maps is confirmed by edge-level speed correlation. The simulations show that our model is able to grasp important structural properties of real urban traffic, as confirmed by a coherent trend in the spatial correlation behavior of congested edges and the associated critical exponent, with respect to traffic volume. 
Encouraged by these results, we finally applied percolation analysis to real and synthetic traffic by comparing the Fisher exponent values, at different times, associated with network fragmentation under load, finding very good agreement. Synthetic experiments were carried out on a variety of different road networks for several kinds of cities. 

Despite not trying to accurately simulate vehicular traffic in the urban context, and explicitly choosing a very coarse vehicular model, the result is a usable tool to quickly compare different city organizations both for testing theoretical ideas and for getting useful glimpses of the main breakup modes of urban networks. Also note that, even though we decided to apply our method to predict congestion patterns typical of urban vehicular traffic, it is expected that other transport phenomena involving agent competition for network resources could be approached in a similar way. In particular, we expect that analyses where BC has provided important insights, such as those on 
power grids, the internet backbone, air travel, and maritime cargo shipping~\cite{barthelemy2011spatial,holme2003congestion,guimera2005worldwide,kazerani2009can}, might benefit from our refined approach. 

\paragraph*{Acknowledgments --} 
We are deeply indebted to Enrico Gobbetti and Francesco Versaci for the many useful discussions. We acknowledge the contribution of Sardinian Regional Authorities under project XDATA (art 9 L.R. 20/2015). Map data copyrighted OpenStreetMap contributors: www.openstreetmap.org

\bibliographystyle{unsrt}
\bibliography{PRE2023}

\end{document}